\documentclass[sigconf]{acmart}
\usepackage{graphicx}

\AtBeginDocument{%
  }

\setcopyright{acmcopyright}
\usepackage{hyperref}
\usepackage{fancyhdr}
\usepackage{tikz}
\usepackage{tcolorbox}

\copyrightyear{2018}
\acmYear{2018}
\acmDOI{XXXXXXX.XXXXXXX}

\begin{document}

\title{TimeLess: A Vision for the Next Generation of Software Development}


\author{Zeeshan Rasheed, Malik Abdul Sami, Jussi Rasku, Kai-Kristian Kemell, Zheying Zhang, Janne Harjamäki, Shahbaz Siddeeq, Sami Lahti, Tomas Herda, Mikko Nurminen, Niklas Lavesson, José Siqueira de Cerqueira, Toufique Hasan, Ayman Khan, Mahade Hasan, Mika Saari, Petri Rantanen, Jari Soini and Pekka Abrahamsson}
\affiliation{
  \institution{Faculty of Information Technology and Communication Science, Tampere University}
  \city{Tampere}
  \country{Finland}
}
\email{zeeshan.rasheed@tuni.fi}
\email{pekka.abrahamsson@tuni.fi}







\pagestyle{fancy}
\fancyhf{} 
\fancyfoot[C]{\thepage} 
\renewcommand{\headrulewidth}{0pt} 

\begin{abstract}

Present-day software development faces three major challenges: complexity, time consumption, and high costs. Developing large software systems often requires battalions of teams and considerable time for meetings, which end without any action, resulting in unproductive cycles, delayed progress, and increased cost. What if, instead of large meetings with no immediate results, the software product is completed by the end of the meeting?
In response, we present a vision for a system called \textbf{TimeLess}, designed to reshape the software development process by enabling immediate action during meetings. The goal is to shift meetings from planning discussions to productive, action-oriented sessions. This approach minimizes the time and effort required for development, allowing teams to focus on critical decision-making while AI agents execute development tasks based on the meeting discussions. We will employ multiple AI agents that work collaboratively to capture human discussions and execute development tasks in real time. This represents a step toward next-generation software development environments, where human expertise drives strategy and AI accelerates task execution.

\end{abstract}


\maketitle

\section{Introduction}
\label{Introduction}

%

Software development is a complex, time-consuming, and expensive process \cite{boehm2000software}. In practice, developers and stakeholders collaborate using established software process models, such as Waterfall, Agile, and DevOps, among others, which are designed to facilitate communication, coordination, and simplify workflows \cite{pargaonkar2023comprehensive}. These processes highlight the inherent complexity, time consumption, and communication challenges that characterize traditional software development \cite{palle2020compare}. The processes require extensive coordination among multiple teams and stakeholders. For instance, the development of large software systems requires teams to hold multiple meetings to discuss strategies, assign tasks, and monitor progress \cite{gaborov2023conceptual}.
These meetings often conclude without generating clear actionable items, resulting in unproductive cycles where discussions fail to translate into immediate progress, ultimately delaying development and increasing costs \cite{kamei2017benefits}. 

\begin{figure}[t]
    \centering
    \includegraphics[width=0.9\linewidth]{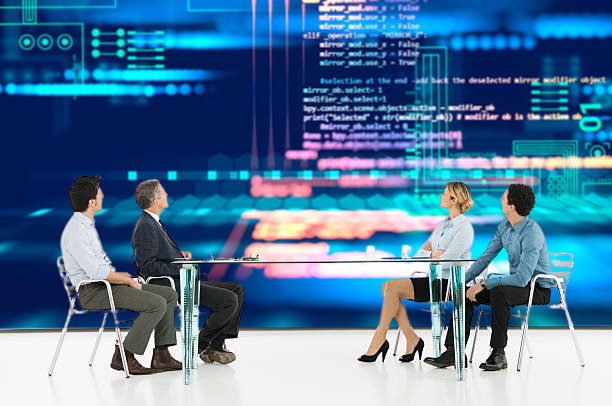}
    \caption{Future Generation of Software Development}
    \label{fig:enter-label}
\end{figure}

In this paper, we present a vision for the next generation of software development, called \textbf{TimeLess}. The aim is to reshape the development process by introducing immediate action and real-time execution during meetings. This approach enables the development of large-scale software projects within a limited time frame by integrating AI, while keeping human teams central to decision-making to drive the process. Meetings shift from planning discussions to action-driven sessions, where teams see the system implement their ideas in real-time. The AI agent will act as an assistant, executing tasks based on team discussions during the meeting, while human developers monitor the agent’s performance to ensure quality and accuracy by providing feedback.



The core concept of \textbf{TimeLess} is to facilitate faster and more efficient software development by allowing AI agents to act as assistants during meetings. 
As shown in Figure \ref{fig:enter-label}, the background screen displays the task execution in real-time, illustrating how AI agents translate discussions among four human stakeholders into actionable development steps.
The process will begin with discussions among stakeholders, developers, architects, and quality assurance personnel regarding the software product. The propose system will listen and capture these discussions, transcribing them into text. The system will process this transcribed data to automatically generate summaries, user stories, epics, and tasks, which will form the foundation for the next stages of development. This iterative process will continue until the team is satisfied with the generated user stories and epics. 
The system will progress through design, coding, testing, and deployment, guided by the human team based on initial user stories generated by multi-AI agents.
Throughout the project, the team will have the flexibility to revisit and update any stage of software development as needed. In this process, the human team will drive the strategy, while AI agents will act as assistants to execute tasks in real time during meetings. Our objective is to make meetings more interactive, productive, and action-oriented, while also reducing the time, cost, and complexity of development.

The introduction of the \textbf{TimeLess} system represents a step forward towards next-generation software development environments that adapt to the demands of the industry. This system aims to enhance productivity by integrating AI-driven processes that facilitate task execution and decision-making. To illustrate this vision, we present initial results in Section \ref{preliminary results}, demonstrating progress toward achieving this goal.

\section{Trends}
\label{Trends}

\begin{figure}[h]
    \centering
    \begin{tikzpicture}
        \matrix[column sep=0.4cm] {
            \node[rectangle, draw, align=center] (adhoc) {Ad hoc}; &
            \node[rectangle, draw, align=center] (waterfall) {Waterfall}; &
            \node[rectangle, draw, align=center] (plan) {Plan driven}; &
            \node[rectangle, draw, align=center] (agile) {Agile}; &
            \node[rectangle, draw, align=center] (TimeLess) {TimeLess}; \\
        };

        \draw[->] (adhoc) -- (waterfall);
        \draw[->] (waterfall) -- (plan);
        \draw[->] (plan) -- (agile);
        \draw[->] (agile) -- (TimeLess);
    \end{tikzpicture}
    \caption{Trends in Software Engineering}
    \label{fig:swe_trends}
\end{figure}
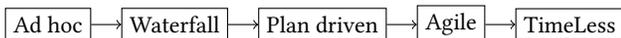

When we examine the development of Software Engineering (SE) as a field separate from computer science, it progresses through several distinct phases, as shown in Figure \ref{fig:swe_trends}). The transitions were often necessitated by the observations and pain points arising from the industry, but also advancements in computing technology has played a major part \cite{pelluru2023advancing}. Rising expectations and capabilities has increased the software complexity, creating the need to evolve SE practices \cite{al2021empirical}.

Initially, software development was done largely in \emph{ad hoc} fashion without formal methodologies guiding the process \cite{cao2007agile}. Management of building bigger and more complex software made these inefficiencies apparent, and the \emph{waterfall} model \cite{saravanos2023simulating} emerged. 
The Waterfall model prioritized planning, but its rigidity and requirement to up front specifications often resulted in overhead and difficulties to adapting to changes. To address these shortcomings, structured or \emph{plan-driven} methodologies \cite{marinho2019plan} were proposed. These approaches introduced more flexibility but still maintained a strong emphasis on upfront planning and formal documentation \cite{wellington2005comparison}. However, they often struggled with real-time adjustments during the development cycle.

The need for more adaptability gave rise to \emph{Agile} methodologies in the early 2000s, which shifted the focus toward iterative development, continuous feedback, and rapid delivery \cite{dingsoyr2012decade}. Agile marked a fundamental change in how time was considered: instead of trying to make intricate plans on sprints and iterations became the units of progress \cite{srivastava2017scrum}. This approach was developed in response to the limitations of previous models, which frequently resulted in a misalignment between the product vision and its final implementation \cite{10176168}. However, despite the increased engagement in sprints and meetings, discussions often fail to produce immediate progress, leading to time-consuming sessions without clear actionable outcomes \cite{salinas2018concerns}.
Recent advances suggest we are entering a new era beyond Agile. The introduction of AI and LLMs are in software development techniques which is changing the SE landscape \cite{10273784}, \cite{hou2023large}. This raises the question of what comes next? 
We feel that this new era needs a name. We propose \textbf{TimeLess} system, because in our vision (elaborated on in Sec \ref{future vision}) many of the time constraints, that necessitated much of the existing processes, are lifted. This makes the feedback cycles shorter. With the the advancement of computing \cite{belzner2023large}, we will move towards even more \textbf{unconstrained} SE practices. For instance, what if meetings end with immediate results with real-time execution with the help of AI?
Another driver in this shift is the recognition that many current SE processes, such as Agile’s ceremonies, are necessary only because of human limitations—context switching, motivation issues, and the lengthy training required for proficiency. The future may involve fewer manual steps, as more decisions and actions are delegated to AI systems that continuously optimize and adjust development processes in real time.

\section{Vision for the Future}
\label{future vision}




\begin{figure}
    \centering
    \includegraphics[width=0.9\linewidth]{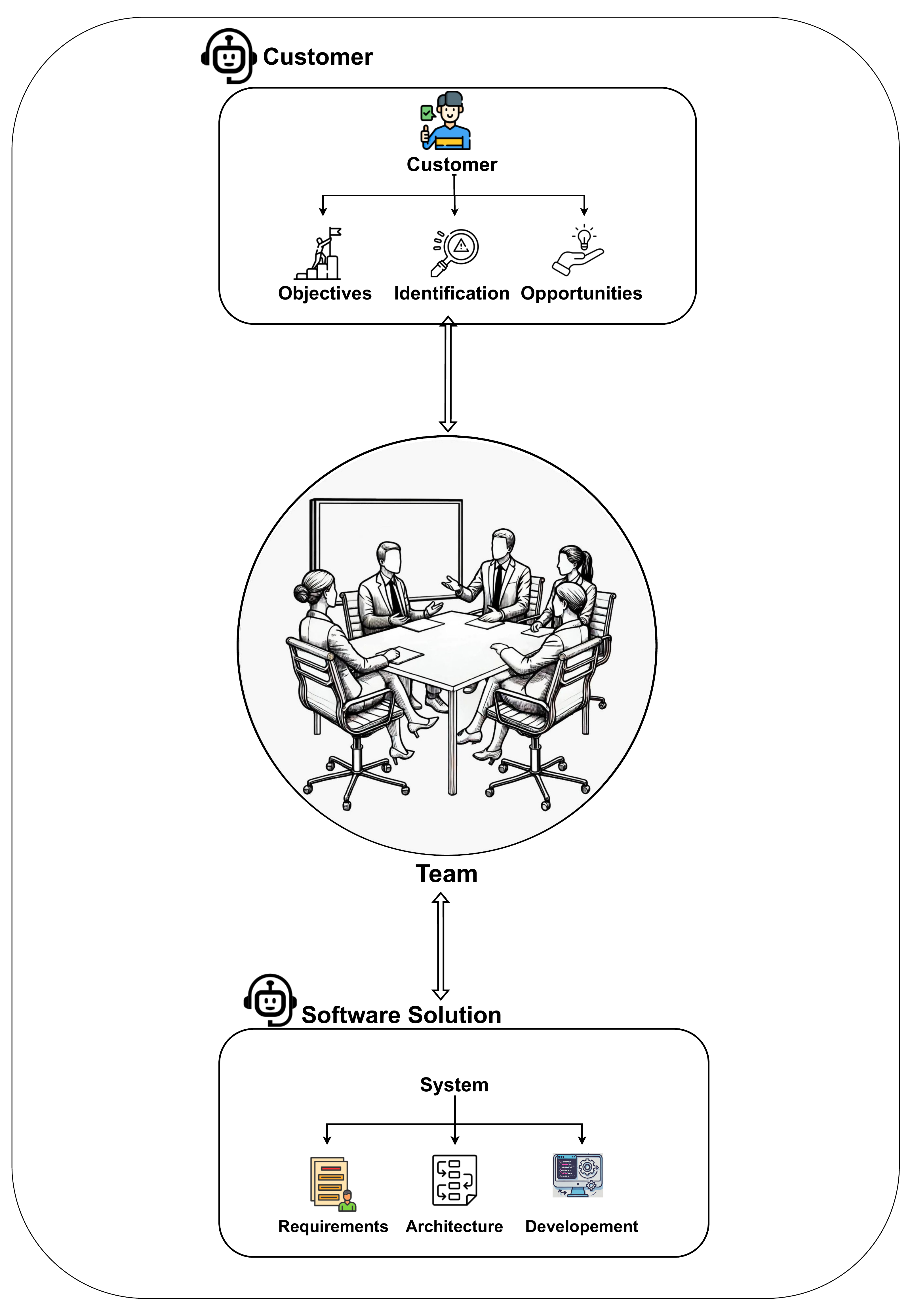}
    \caption{TimeLess System Workflow}
    \label{fig:unified_framework}
\end{figure}



Despite the trends and evolution in SE practices, a challenge remains: how can we build and maintain control over increasingly complex software systems while considering the limited time and other resources? Discussion on trends like \emph{AI-driven development} \cite{ernst2022ai} provide insights into future developments, but a more concrete vision is needed to guide our collective efforts.

Our vision is a software development environment where the team and customer interact with the AI system, with meetings focused on meaningful goals and processes driven by humans. This setup raises the level of abstraction from code elements to core software concepts, such as user stories, features, and user experience. By rethinking the software development process, we establish a system where AI assists in creating applications shaped by both teams and customers. This approach broadens access to SE while maintaining standards of quality and accuracy. The key realization in the \textbf{TimeLess} vision is that AI reduces time constraints in software development. Team and customer interactions often follow structured processes, but meetings and procedural issues take time and cause delays. \textbf{TimeLess} uses AI to shorten traditional sprint duration's from weeks to minutes, allowing for real-time software generation during collaborative meetings. This approach enables teams to focus on immediate and relevant outcomes.


The \textbf{TimeLess} system redefines the interaction between developers and integrated development environments by incorporating autonomy in executing various agile practices. As shown in Figure \ref{fig:unified_framework}, It supports the development team and stakeholders by adapting to inputs from meetings, enhancing communication across project phases, and managing expectations through immediate outputs. Acting almost as a participant, the system interprets conversational input and translates it into technical specifications and implementations. Our approach centers on the role of the development team and customer, ensuring the system responds to collaborative inputs, reducing project risks, and enhancing software quality. Real-time generation of software components addresses miscommunication while upholding quality and accuracy, combining the structured rigor of waterfall methodologies with the flexibility of agile practices. By providing immediate visual feedback and technical specifications, the system offers all stakeholders a clear and consistent understanding of project progress and expectations.

The \textbf{TimeLess} system engages both technical team members and non-technical stakeholders through design and implementation phases, addressing technical questions as they arise. It manages the complete software lifecycle, including testing, integration, validation, and deployment, by using established SE principles to address lifecycle considerations effectively. To realize this vision, we recognize we are entering new territory. However, with advancements in AI, it becomes possible to integrate the rigor of waterfall methodologies with the quick feedback cycles of agile. Our ultimate goal is to develop a concrete platform to implement the \textbf{TimeLess} approach, which requires further research and methodological innovation.

The following sections outline the different aspects of our vision, including user requirement gathering, communication with both customer and team, and lifecycle considerations for the created software. We discuss the feasibility and technical implementation details later in Sections \ref{proposed system} and \label{preliminary results}, where we present the \textbf{TimeLess} system architecture and preliminary results from experiments on key technologies.

\section{Proposed System}
\label{proposed system}

\begin{figure}
    \centering
    \includegraphics[width=0.9\linewidth]{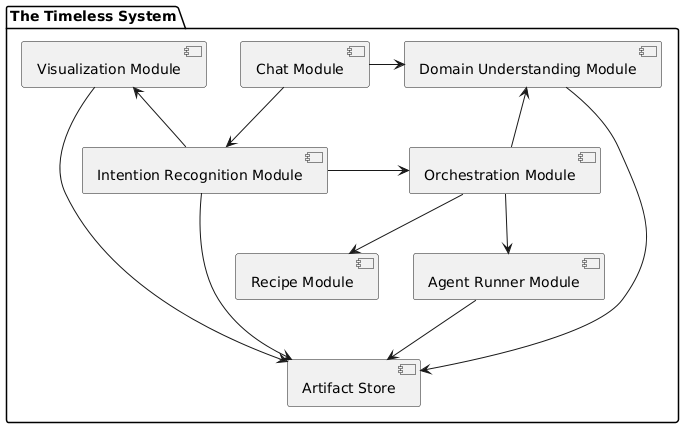}
    \caption{Components of the Proposed TimeLess System}
    \label{fig:proposed_module}
\end{figure}

In this section, we discuss the details on how to realize the vision in the form of an AI-assisted development environment. The system has a modular structure (see Figure~\ref{fig:proposed_module}), with each module serving a different purpose.

The \textbf{\textit{Chat Module}} listens to the team conversation, performs speaker identification, and produces transcripts. It incorporates a voice-based user interface with Text-to-Speech (TTS), Voice Transformation Technology (VTT), and speaker recognition \cite{basit2024tinydigiclones}. This enables real-time communication and interaction between users and the system. The \textbf{\textit{Visualization Module}} informs the team of the system state and progress, displaying artifacts and outputs when requested. This module provides visual representations of team meetings, project discussions, progress metrics, processes, and ongoing activities, facilitating an overview of the project's status. 

The \textbf{\textit{Domain Understanding Module}} reads external materials and project files, follows discussions, and builds and improves the understanding of the project context. The module relies on LLMs and Retrieval-Augmented Generation (RAG) techniques \cite{jiang2023active} to extract relevant information from the web and other sources, generating prompts for other modules to ensure accurate, context-aware responses and actions for error-free software development \cite{lewis2020retrieval}. The \textbf{\textit{Intention Recognition Module}} interprets the team discussion transcripts, maintains the system state, updates current goals, and detects when consensus has been reached and when the team is ready to move on. This analyzes user goals during meetings and prepares an initial orchestration setup to configure the interface and architecture based on user intentions. It enhances task coordination across software development workflows with the help of specialized agents \cite{jiang2023active}.

The \textbf{\textit{Orchestration Module}} allocates resources, sets agent roles, and builds prompts. It manages resource allocation, agent spawning, capabilities, and workflow execution, ensuring that tasks are efficiently distributed and managed across the system. The \textbf{\textit{Recipe Module}} contains the Software Engineering Body of Knowledge (SWEBOK) and best practices. It is used to find templates on how to work, providing standardized procedures and guidelines to ensure consistency and quality in the development process.

The \textbf{\textit{Agent Runner Module}} manages agents based on tasks set by the orchestration module, interprets their output, and stores it. It ensures that each agent operates within its role and that their contributions are integrated into the project. Finally, the \textbf{\textit{Artifact Store}} serves as a repository for all project-related documents and specifications once all software requirements and specifications are completed. It stores artifacts such as user stories, Non-functional Requirements (NFRs), UI mockups, unit tests, API documentation, release notes, and user interfaces, ensuring that all project materials are securely and systematically organized.

\section{Preliminary Results}
\label{preliminary results}
In this section, we present preliminary results that represent a step forward toward achieving our vision. The main aim is to explore the feasibility of AI agents in minimizing the time and effort required for large-scale software development. Our initial results show that the multi-agent system has the ability to accomplish the vision of the \textbf{TimeLess} system (elaborated in Sec \ref{future vision}).

In our initial experiment \cite{sami2024experimenting}, we proposed a platform that converts user requirements into structured software development outputs such as user stories, prioritization, UML diagrams, front-end code, back-end code, unit tests, and end-to-end tests. Initially, the proposed platform generate user stories based on given requirements and applied prioritization techniques such as the Analytic Hierarchy Process (AHP), the 100 Dollar Test, and Weighted Shortest Job First (WSJF). The goal was to automate the generation and prioritization of requirements by integrating LLM based agents. We tasked the multi-agent system with converting these requirements into user stories and epics, then prioritizing them using AHP, 100 dollor technique, and WSJF methods. These methods ranked the user stories according to importance, producing results within seconds.These methods ranked the user stories based on their importance to the requirements. Our findings show that an LLM-based multi-agent system has potential for automating requirement prioritization, advancing the goals of the \textbf{TimeLess} project environment.  

The next step is to generate a UML diagram from the user requirements. The system converts user input into the specific textual format required by PlantUML, known as the PlantUML response, and processes this data through an API call to create visual UML representations. These diagrams are then displayed on the client-side application using the MIME type svg/xml.After generating the UML diagram, the next phase is automating code generation for back-end development, front-end development, back-end unit testing, and client-side test case generation using an LLM-based multi-agent system. Our contribution includes dynamically generating code for each of these areas, speeding up both the development and testing processes.

The results show the proposed platform generates code. However, when it comes to large-scale projects, the system fails. To address this, we propose a multi-agent system that autonomously generates code for large and complex projects by providing high-level descriptions as input \cite{rasheed2024codepori}. Our proposed system uses a multi-agent system, where each agent is designed to specialize in different aspects of software development, from understanding requirements to writing and optimizing code, thus taking on different roles in the collaboration process. The proposed system is capable of generating between 1,500 and 2,000 lines of code. We publicly released a dataset that can help researchers and practitioners access all the collected data for validating our study \cite{replpack}. As presented in Table \ref{Result Codepori}, we provide the results for the five input projects, detailing the respective lines of code and the time required to complete each task. We also validates our proposed system with recently developed models for code generation field. The results indicate that the proposed system improve code accuracy of 89\%. We also set up the docker environment, utilizing multiple AI agents to configure docker and execute the code within it. These AI agents handled the docker setup, ensuring all dependencies and settings were properly applied, allowing the program to run in an isolated environment \cite{Honkanen2024}.

\begin{table}[]
\caption{Result produced by CodePori}
\label{Result Codepori}
\begin{tabular}{|l|l|l|l|l|l|}
\hline
\textbf{S.No} & \textbf{Input ID} & \multicolumn{1}{c|}{\textbf{\begin{tabular}[c]{@{}c@{}}Line of\\ Codes\end{tabular}}} & \multicolumn{1}{c|}{\textbf{Mins}} & \textbf{Modules} & \textbf{OpenAI Bill} \\ \hline
01            & D1                & 477                                                                                   & 20                                 & 5                & 1.45\$               \\ \hline
02            & D5                & 444                                                                                   & 18                                 & 2                & 1.20\$               \\ \hline
03            & D4                & 580                                                                                   & 27                                 & 4                & 1.80\$               \\ \hline
04            & D6                & 789                                                                                   & 35                                 & 7                & 2.10\$               \\ \hline
05            & D7                & 1180                                                                                  & 40                                 & 9                & 2.56\$               \\ \hline
\end{tabular}
\end{table}

To improve accuracy and reduce hallucination, we developed a RAG-based system that combines information retrieval with natural language generation techniques to produce more accurate, contextually grounded responses. Our system retrieves and generates answers based on uploaded content, enabling real-time, context-aware interactions with project-related content \cite{GPTLaboratory2024}. By integrating the RAG system into the \textbf{TimeLess }system, users can query and receive precise answers regarding requirements, design constraints, or existing code, facilitating efficient decision-making and reducing hallucination.

\section{Conclusions}
\label{conclusions}
In this paper, we present our vision, called the \textbf{TimeLess} system, which shifts the focus from lengthy and unproductive meetings to action-oriented sessions, where the proposed system enables immediate task execution with the support of AI agents. This reduces unproductive cycles and accelerates progress, allowing teams to complete software tasks more efficiently. By experimenting with our developed system (detailed in Section \ref{preliminary results}), we have gained insights into how AI technology can fundamentally reshape software development and pave the way for future generations of SE.

Our experiments with the system, as detailed in Section \ref{preliminary results}, demonstrate its applications and how SE is approached. The findings show that AI agents can support human tasks, automate routine processes, and improve decision-making. These results indicate that such technology enhance current practices and lay the foundation for a new phase in SE, where AI integration becomes central to development processes. This vision offers an exciting pathway for future research, opening new avenues for exploring AI’s role in enhancing collaboration, productivity, and the scalability of software projects. The \textbf{TimeLess} system represents a key step toward realizing the full potential of AI in reshaping SE for future generations.


\bibliographystyle{ACM-Reference-Format}
\bibliography{acmart} 

\end{document}